\documentclass{PoS}

\usepackage{amssymb}
\usepackage{fontenc}
\usepackage{times}
\usepackage{mathptmx}
\usepackage{color}
\usepackage{graphicx}
\usepackage{ifpdf}

\usepackage{psfrag}
\usepackage{bm}  

\PoS{PoS(LAT2005)286}

\title{The IR gluon propagator from lattice QCD}

\ShortTitle{The IR gluon propagator from lattice QCD }

\author{\speaker{Paulo J. Silva}\thanks{Supported by FCT via grant 
SFRH/BD/10740/2002.} ~and  Orlando Oliveira \\
Centro de Física Computacional, Departamento de Física, 
Universidade de Coimbra     \\
3004-516 Coimbra, Portugal. \\
E-mail: \email{psilva@teor.fis.uc.pt}, \email{orlando@teor.fis.uc.pt}}

\abstract{The gluon propagator is computed in large asymmetric lattices, 
 accessing momenta around 100 MeV and smaller. Our study tries to 
 check the compatibility of the recent solutions of the gluon-ghost 
 Dyson-Schwinger equations with lattice results. In particular the 
 exponent $\kappa$, which characterizes the solutions for the infrared, is 
 measured. Results favours a vanishing zero momentum gluon propagator. We also 
 report on the compatibility of various functional forms used to fit 
 the Dyson-Schwinger solution for the full range of momenta and the 
 lattice gluon propagator.}

\FullConference{XXIIIrd International Symposium on Lattice Field Theory\\
                 25-30 July 2005\\
                 Trinity College, Dublin, Ireland}

\begin{document}

%\section{Introduction and motivation}

Quantum chromodynamics (QCD) is the theory that describes the interaction 
between quarks and gluons. Two first principles approaches to non-perturbative
QCD are Dyson-Schwinger equations (DSE) and the lattice formulation of QCD
(LQCD). Solving the DSE, an infinite tower of equations, requires a truncation
scheme and a parametrization of higher order vertices \cite{AlSm01}. On the 
other hand, in LQCD one has to care about finite volume \cite{OlSi05} and 
finite space effects. If handling finite space effects requires smaller 
lattice spacing, i.e. higher $\beta$ values, finite volume effects requires 
large volumes. 

Recently, there has been some progress on the solution of the DSE. Assuming
ghost dominance, the equations were solved in the infrared (IR) limit. The 
solution shows a vanishing zero momentum gluon propagator \cite{LeSm02}. 
Moreover, the 
coupled gluon-ghost equations were solved numerically and analitical
functions found that fitted the solution \cite{Al04,Fi03}. 
On the lattice, the gluon propagator has been revisited a number of times. 
However, due to the lattice sizes used in the investigations, the IR region 
was not yet properly accessed. The motivation of our study is to look at the 
deep IR gluon propagator and to try
to check the compatibility between the two approaches, i.e. to test the IR 
DSE solution for the gluon propagator and to see if the same functional forms 
describe the lattice data and the DSE solution. In order to try to investigate
the deep IR region, we consider large assymetric 4D lattices, i.e. lattices 
with a very large time extension \cite{OlSi04}.

\section{The lattice setup}

In this investigation we consider SU(3) pure gauge, Wilson action at 
$\beta = 6.0$. The value of the lattice spacing to convert lattice into 
physical units being $a^{-1} = 1.943$ GeV \cite{Bali}. For the lattices
$16^3 \times 128$ and $16^3 \times 256$, the gauge configurations\footnote{All
configurations were generated with MILC code 
{\tt http://physics.indiana.edu/\~{ }sg/milc.html}.} 
were generated as described
in \cite{OlSi04}. For the smallest (largest) lattice 164 (155) configurations
were generated. The large time extension allows to
access momenta of about 95 MeV for the smallest lattice and 48 MeV for 
the largest lattice. For the definition of the gluon propagator, notation and 
the gauge fixing method see \cite{gribovgluon}.

\section{Gluon dressing function}

\begin{figure}[h]
\psfrag{EIXOX}{{\huge $q(GeV)$}}
\psfrag{EIXOY}{{\huge $q^2 D(q^2)$}}
%\subfigure[$16^3\times128$]{ 
   \begin{minipage}[b]{0.45\textwidth}
   \centering
   \includegraphics[angle=-90,origin=c,scale=0.27]{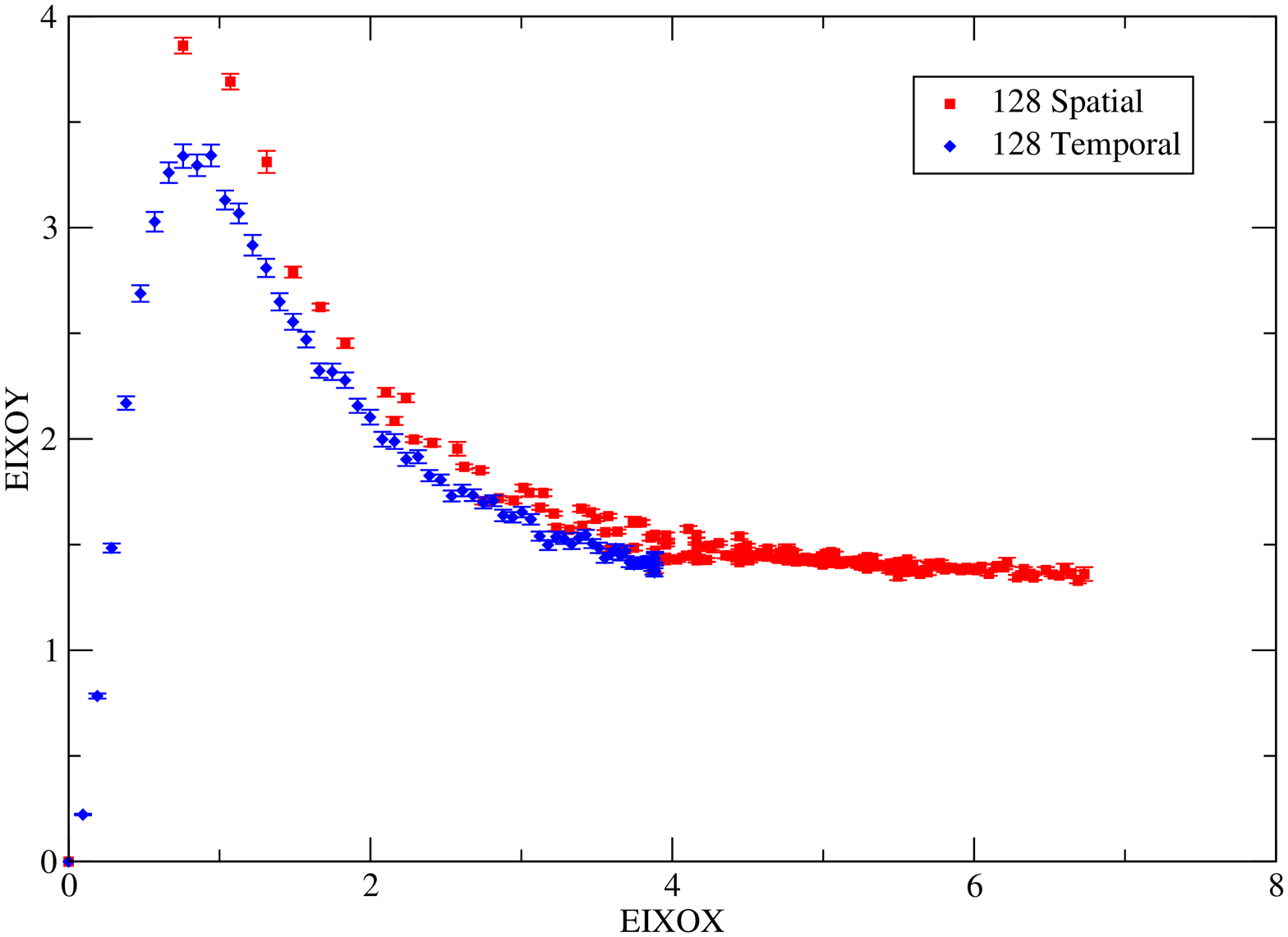}
   \end{minipage} 
%}
%\subfigure[$16^3\times256$]{ 
  \begin{minipage}[b]{0.45\textwidth}
  \centering
  \includegraphics[angle=-90,origin=c,scale=0.27]{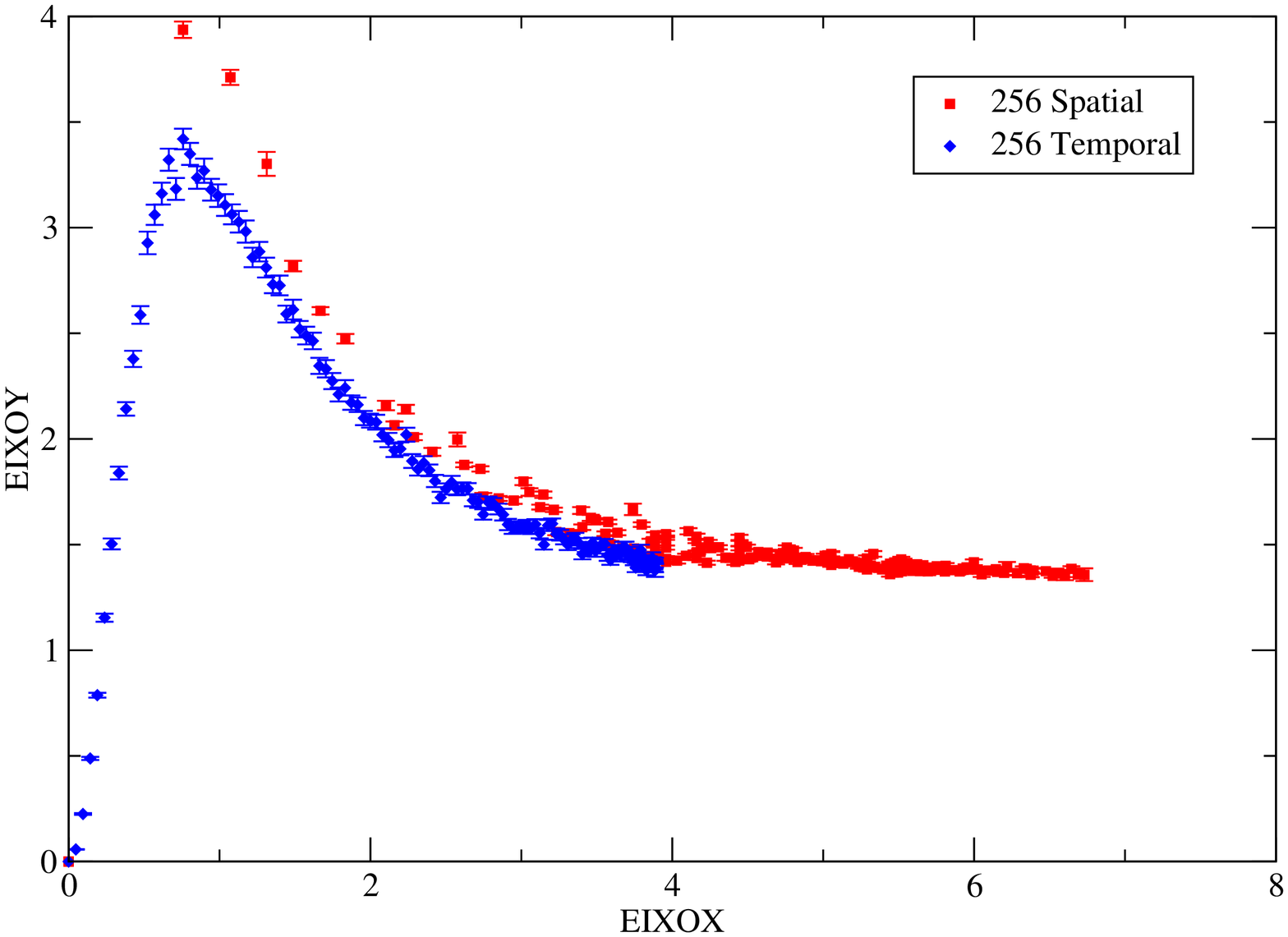}
  \end{minipage} 
%}
\vspace*{-1.5cm}
\caption{Gluon dressing function for the smaller (left) and
larger (right) lattices.}
\label{Zallmomenta}
\end{figure}

In figure \ref{Zallmomenta} the bare gluon dressing function is shown for
all pure temporal and pure spatial momenta. The data shows volume 
effects, with
the spatial momenta giving larger values for the dressing function. The data
suggests that one should not mix temporal and spatial momenta.

\begin{figure}[h]
\psfrag{EIXOX}{{\huge $q(GeV)$}}
\psfrag{EIXOY}{{\huge $q^2 D(q^2)$}}
%  \subfigure[Dressing function]{ 
   \begin{minipage}[b]{0.45\textwidth}
   \centering
   \includegraphics[angle=-90,origin=c,scale=0.27]{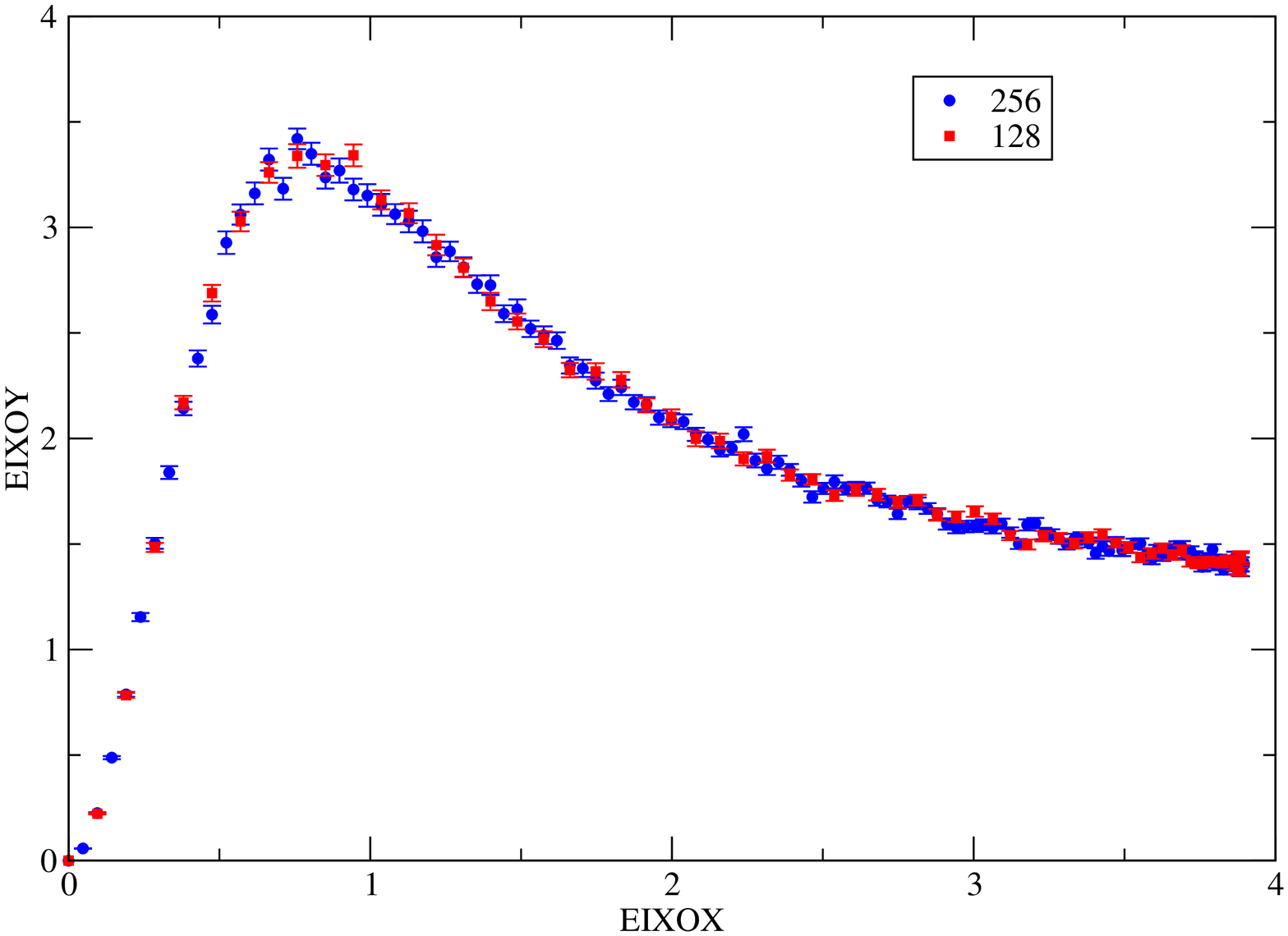}
  \end{minipage} 
% }
\psfrag{EIXOY}{{\huge $D(q^2)$}}
%\subfigure[Raw propagator]{ 
  \begin{minipage}[b]{0.45\textwidth}
  \centering
  \includegraphics[angle=-90,origin=c,scale=0.27]{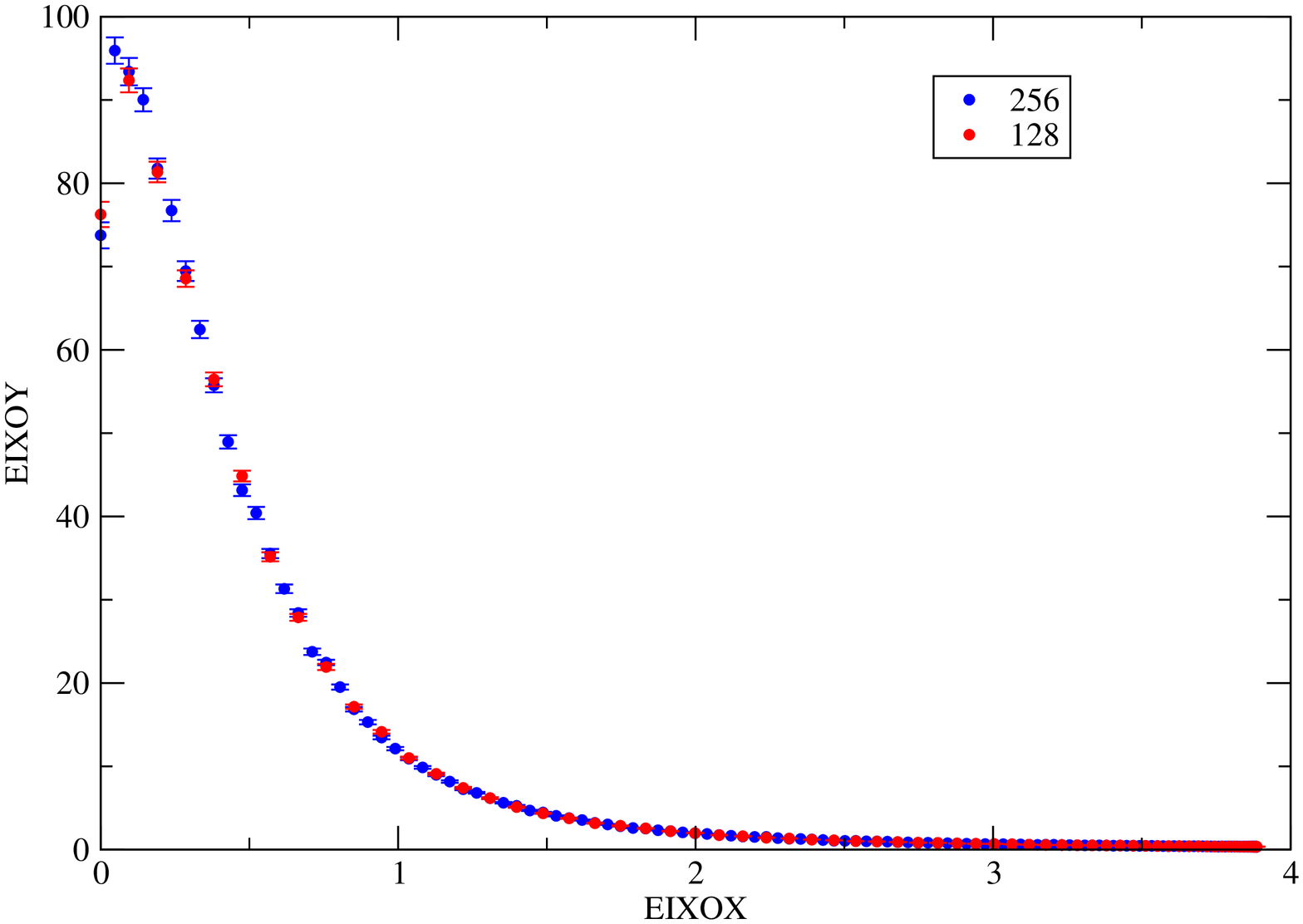}
  \end{minipage} 
%}
\vspace*{-1.5cm}
\caption{Gluon dressing function (left) and propagator (right) for pure 
         temporal momenta.} 
\label{GZD}
\end{figure}

For the IR study, we will consider only pure temporal momenta. For these 
momenta and the two lattices, the gluon dressing function and the propagator 
are reported in figure \ref{GZD}. The two data sets are compatible within 
errors. Note that this does not mean that finite volume effects are 
negligable \cite{OlSi05}.

\subsection{IR gluon dressing function}

The IR analytical solution of the DSE for the gluon dressing function being
$Z(q^2)\sim(q^{2})^{2\kappa}$, we first try to fit the lattice data to this 
expression\footnote{The fits were done separately for each lattice. The
statistical errors on all fitting parameters were estimated using 1600
bootstrap samples.}. An acceptable $\chi^2/d.o.f. = 0.40$ was obtained only by
fitting the first three lowest momenta (excluding the zero momentum) for the 
largest lattice ($| q | \le 143$ MeV), meaning that the DSE solution seems to
be valid only for momenta below 150 MeV. The measured 
$\kappa=0.4859^{+22}_{-24}$ does not support a vanishing zero momentum gluon 
propagator. In order to test the validity of the result, corrections to the 
DSE solution were considered, fitting the smallest possible range of momenta.
The results are in table 1.

\begin{table}[h]
\begin{center}
\begin{tabular}{|lccc|}
\hline
$16^3\times 256$ & $\bm{q_{max}}$ & $\bm{\kappa}$
  & \textbf{$\chi^2 /d.o.f.$}  \\
\hline
$\omega(q^2)^{2\kappa}(1+aq^2)$ & 
               $ 191 MeV$ & $0.5070^{+36}_{-50}$  &   $0.44$ \\
$\omega(q^2)^{2\kappa}(1+aq^2+bq^4)$ & 
               $ 238 MeV$ & $0.5131^{+67}_{-64}$  &   $1.03$ \\
$\omega(q^2)^{2\kappa}(1+aq^2+bq^4+cq^6)$ & 
               $286 MeV$ & $0.5146^{+70}_{-96}$ &  $1.36$ \\
\hline
\end{tabular}
\caption{Fits to polinomial corrections to the DSE analytical solution.}
\label{irDSEc}
\end{center}
\end{table}

It is interesting to observe that, with corrections to the analitycal DSE 
solution, $\kappa$ becomes larger than 0.5. Moreover, the results reported in
table 1 are compatible within errors and give an error weighted 
mean value of $\kappa = 0.5108(13)$. Note that the $\kappa$ just reported are 
not compatible with the direct fit to the DSE solution.

In \cite{Al04} the DSE solution was well described by the two expressions
\begin{equation}
   Z_{cut}(q^2)=\omega\left(\frac{q^2}{\Lambda_{QCD}^2+q^2}\right)^{2\kappa} 
   \quad , \hspace{-0.2cm}\quad
   Z_{pole}(q^2)=\omega\frac{\left(q^2\right)^{2\kappa}}
                            {\left(\Lambda_{QCD}^2\right)^{2\kappa}+
                             \left(q^2\right)^{2\kappa}} \quad ,
\end{equation}
which associate, respectively, a cut and a pole to the propagator in the IR
region. The lattice data is well described by both expressions with no clear
preference for the cut or the pole. The results of the fits, for the momenta 
region starting at the lowest nonzero momentum up to $q_{max}$, are reported
in table \ref{ir2}. Note that all fits have good $\chi^2/d.o.f.$ and, for each
function, the fitted parameters are compatible within one standard deviation.

\begin{table}[h]
\begin{center}
\begin{tabular}{|cccccc|}
\hline
           & Lattice           & 
             $q_{max}$         & 
             \textbf{$\kappa$} & 
             \textbf{$\Lambda_{QCD}$} (MeV)   & 
             \textbf{$\chi^2 /d.o.f.$}  \\
\hline
$Z_{cut}$ & $16^3\times 256$     & 
            $ 664 MeV$           & 
            $0.5090^{+19}_{-20}$ & 
            $409^{+4}_{-4}$      &  
            $0.71$               \\
          & $16^3\times 128$     & 
            $570 MeV$            & 
            $0.5117^{+48}_{-46}$ & 
            $417^{+8}_{-8}$      &  
            $1.25$ \\
\hline
$Z_{pole}$ & $16^3\times 256$     & 
             $664 MeV$            & 
             $0.5077^{+16}_{-17}$ & 
             $409^{+4}_{-3}$      &  
             $0.69$               \\
           & $16^3\times 128$     & 
             $570 MeV$            & 
             $0.5100^{+38}_{-31}$ & 
             $416^{+6}_{-8}$      &  
             $1.15$               \\
\hline
\end{tabular}
\caption{$Z_{cut}$ and $Z_{pole}$ fits.}
\label{ir2}
\end{center}
\end{table}

\subsection{UV gluon dressing function}

As in previous studies \cite{leinweber,gribovgluon} we fit the UV data
to the one-loop perturbative gluon dressing function,
\begin{equation}
   \omega \, \left[ \frac{1}{2} \ln \left(\frac{q^2}{\Lambda^2}
                                    \right)
             \right]^{-\gamma}
\end{equation}
where $\gamma = -13/22$ is the gluon anomalous dimension. In the UV region two
types of momenta were considered, namely pure temporal and cylinder plus 
conic cuts as defined in \cite{leinweber}. The results can be seen in 
table \ref{uvfit} for the largest acceptable momentum range.
They show that the UV region is well described by the one-loop perturbative 
result, with the temporal momenta following the perturbative behaviour earlier
than the cylindrical+conical cut momenta. Note that since the fits are to the 
bare data, the fitted $\Lambda$'s should not be compared directly with 
standard values.

\begin{table}[h]
\begin{center}
\begin{tabular}{|l|cc|cc|}
\hline
 \textbf{Lattice} & \multicolumn{2}{c|}{\textbf{Temporal}}  & 
                    \multicolumn{2}{c|}{\textbf{Cuts}}     \\
                  & \multicolumn{2}{c|}{$1.8 GeV \leq\|q\|\leq 3.9 GeV$} & 
                    \multicolumn{2}{c|}{$2.8 GeV \leq\|q\|\leq 7.8 GeV$} \\
                  & $\bm{\Lambda}$ & $\bm{\chi^2 /d.o.f.}$ & 
                    $\bm{\Lambda}$ & $\bm{\chi^2 /d.o.f.}$ \\
\hline
$16^3\times 128$ & $1006^{+7}_{-9} MeV$        &  
                   $1.2$                       &
                   $640^{+5}_{-5} MeV$         &  
                   $1.1$                       \\
$16^3\times 256$ & $990^{+6}_{-6} MeV$         &  
                   $1.1$                       & 
                   $643^{+4}_{-4} MeV$         &  
                   $1.2$ \\
\hline
\end{tabular}
\caption{UV fits.}
\label{uvfit}
\end{center}
\end{table}

\subsection{The gluon dressing function}

In what concerns the full data set, it was tested against the 
following formulas
%\begin{equation}
 $Z_{fit}(q^2) = B(q^2) \left(\alpha_{fit}(q^2)\right)^{-\gamma}\, , $
%\end{equation}
with $B(q^2)=Z_{pole,cut} (q^2)$ and using two different definitions for the 
running coupling, 
\begin{eqnarray}
  \alpha_{P}(q^2) & =  &
    \frac{1}{1+\frac{q^2}{\Lambda_{QCD}^2}} \, 
    \bigg[ \alpha(0) + \frac{q^2}{\Lambda_{QCD}^2}\times 
           \frac{4\pi}{\beta_0}\Big(\frac{1}{\ln(q^2/\Lambda_{QCD}^2)} - 
           \frac{1}{q^2/\Lambda_{QCD}^2-1}\Big)\bigg]\, , 
\, \beta_0=11 ,  \\
\alpha_{LN}(q^2) & = & \frac{\alpha(0)}
                        {\ln\left[e+a_1(q^2/\Lambda_{QCD}^2)^{a_2}\right]}
\end{eqnarray}
proposed in \cite{Al04,Fi03}. The fits with 
$\alpha_{P}(q^2)$ had always $\chi^2/d.o.f. \, \ge \, 2$, i.e. the lattice data
is not described by such a running coupling constant. The lattice data adjusts
better to $\alpha_{LN}(q^2)$ as can be seen by the results reported in
table 4. Note that the $\kappa$ values are now larger and closer
to the DSE figures, than those measured in the IR region.

\begin{table}[h]
\begin{center}
\begin{tabular}{|lccccc|}
\hline
    \textbf{$\bm{B=Z_{cut}}$}
  & $\bm{\kappa}$
  & \textbf{$\bm{\Lambda_{QCD}}$ (MeV)}
  & $\bm{a_1}$
  & $\bm{a_2}$
  & \hspace{-0.2cm}$\bm{\chi^2 /d.o.f.}$  \\
\hline
$16^3\times 128$ & $0.5435^{+36}_{-41}$ &  $364^{+4}_{-4}$ & $0.0062^{+3}_{-3}$ & $2.44^{+2}_{-1}$  & $1.82$ \\
$16^3\times 256$ & $0.5244^{+21}_{-15}$ & $374^{+2}_{-2}$  & $0.0072^{+3}_{-3}$ & $2.424^{+10}_{-11}$ & $1.73$ \\
\hline
\hline
    \textbf{$\bm{B=Z_{pole}}$}
  & $\bm{\kappa}$
  & \textbf{$\bm{\Lambda_{QCD}}$ (MeV)}
  & $\bm{a_1}$
  & $\bm{a_2}$
  & $\bm{\chi^2 /d.o.f.}$  \\
\hline
$16^3\times 128$ & $0.5335^{+23}_{-26}$ &  $373^{+3}_{-3}$ & $0.0081^{+3}_{-3}$ & $2.36^{+2}_{-2}$  & $1.65$ \\
$16^3\times 256$ & $0.5217^{+14}_{-14}$ & $377^{+2}_{-2}$  & $0.0082^{+3}_{-3}$ & $2.36^{+1}_{-1}$ & $1.61$ \\
\hline
\end{tabular}
\end{center}
\label{dse1}
\caption{Fits to all lattice data.}
\end{table}

The value of the zero momentum running coupling can be computed from the 
asymptotic behaviour of the QCD $\beta$-function and from $\alpha_{LN}(p^2)$, 
$\alpha (0) = (4\pi/\beta_0)a_2$. Note that the values reported in table 5
 are not too far from the DSE numbers.

\begin{table}[h]
 \begin{center}
 \begin{tabular}{|ccc|}
\hline
$\bm{\alpha(0)}$ &  $ Z_{cut}$  & $Z_{pole}$  \\
\hline
 $16^3\times128$  &  $2.79^{+2}_{-1}$ & $2.70^{+2}_{-2}$\\
 $16^3\times256$  &  $2.77^{+1}_{-1}$ & $2.70^{+1}_{-1}$\\
 \hline
 \end{tabular}
 \end{center}
\label{alpha0}
\caption{Values of $\alpha(0)$ extracted from $\alpha_{LN}(p^2)$.} 
\end{table}

\section{Gribov copies}

\begin{figure}[h]
\begin{center}
\psfrag{EIXOX}{{\huge $\hat{n}_t$}}
\psfrag{EIXOY}{{\huge $D_{ID}/D_{CEASD}$ }}
\includegraphics[angle=-90,origin=c,scale=0.35]{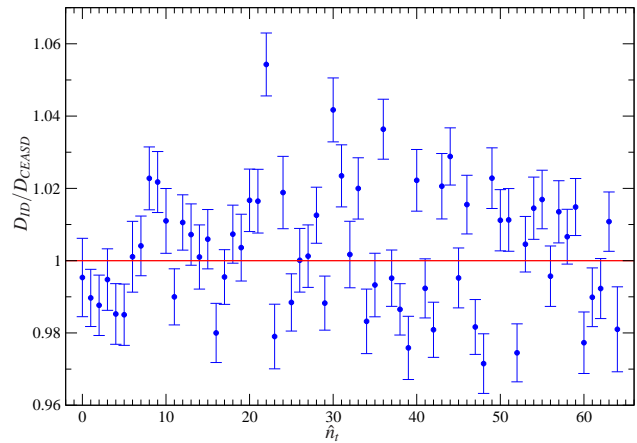}
\vspace*{-1.75cm}
\caption{Gribov copy effects in $16^3\times 128$ configurations.}
\label{gribov}
\end{center}
\end{figure}

For the smallest lattice, we look for possible Gribov copies effects comparing
the method used in \cite{OlSi04} with that described in \cite{ceasd}. The
ratio of the propagators computed from the two methods, $D_{ID}/D_{CEASD}$,
for pure temporal momenta is shown in figure \ref{gribov}. Although we have a 
ratio that, for many momenta, is not compatible with one, we cannot conclude
in favour of any systematic 
effect\footnote{In \cite{gribovgluon} the effect of the Gribov copies was 
estimated as a 2 to $3\sigma$ effect.}. This means either that our statistics 
is too small to resolve Gribov copies or they do not play a significant role 
for such lattices.

\section{Conclusions}

The lattice data seems to be compatible with functional forms that
describe the gluon-ghost DSE numerical solution. In what concerns the vanishing
of the zero momentum gluon propagator, our data does not provide a clear 
answer but it favours a vanishing propagator in agreement with the DSE
solution, flow equations studies \cite{Pa04,FiGi04} 
and time independent stochastic quantisation \cite{Zw03}.


\begin{thebibliography}{99}
\bibitem{AlSm01} See, for example, R. Alkofer, L. von Smekal, 
                 \emph{Phys. Rept.} 
                 {\bf 353} (2001) 281 [{\tt hep-ph/0007355}].
\bibitem{OlSi05} For an exploratory study of the finite volume effects on IR
                 gluon propagator see O. Oliveira, P. J. Silva, 
                 "Finite volume effects in the gluon propagator", 
                  these proceedings [{\tt hep-lat/0509037}].
\bibitem{LeSm02} C. Lerche, L. von Smekal, 
                 \emph{Phys. Rev.} {\bf D65} (2002) 125006 
                 [{\tt hep-ph/0202194}].
\bibitem{Al04} R. Alkofer, W. Detmold, C. S. Fischer, P. Maris, 
               \emph{Phys. Rev.} {\bf D70} (2004) 014014  
                [{\tt hep-ph/0309077}].
\bibitem{Fi03} C. S. Fischer, R. Alkofer, 
               \emph{Phys. Rev.} {\bf D67} (2003) 094020 
               [{\tt hep-ph/0301094}].
\bibitem{OlSi04} O. Oliveira, P. J. Silva, {\emph AIP Conf. Proc.} {\bf 756}
                 (2005) 290 [{\tt hep-lat/0410048}].
\bibitem{Bali} G. S. Bali, K. Schilling, \emph{Phys. Rev.} 
               {\bf D47} (1993) 661 [{\tt hep-lat/9208028}]. 
\bibitem{gribovgluon} P. J. Silva, O. Oliveira, 
                      {\emph Nucl. Phys.} {\bf B690} (2004) 177 
                      [{\tt hep-lat/0403026}].
\bibitem{leinweber} D. B. Leinweber \textit{et al} {\emph Phys. Rev.} 
                    {\bf D60} (1999) 094507; 
                erratum-ibid {\emph Phys. Rev.} {\bf D61} (2000) 079901
                 [{\tt hep-lat/9811027}].
\bibitem{ceasd} O. Oliveira, P. J. Silva, {\emph Comp. Phys. Comm.} 
              {\bf 158} (2004) 73 [{\tt hep-lat/0309184}].
\bibitem{Pa04} J. M. Pawlowski, D. F. Litim, S. Nedelko, L. von Smekal,
              {\emph Phys. Rev. Lett.} {\bf 93} (2004) 152002.
\bibitem{FiGi04}C. S. Fischer, H. Gies, {\emph JHEP} {\bf 0410} (2004) 048
              [{\tt hep-ph/0408089}].
\bibitem{Zw03} D. Zwanziger, {\emph Phys. Rev.} {\bf D67} (2003) 105001
              [{\tt hep-th/0206053}].
\end{thebibliography}
\end{document}